\documentclass[sigconf]{aamas}
\setcopyright{none}
\renewcommand\footnotetextcopyrightpermission[1]{}

\usepackage{balance} 
\usepackage{svg}

\usepackage{fancyhdr}
\fancypagestyle{firstpage}{
    \fancyhf{}

    \fancyfoot[L]{%
        \footnotesize
        \rule{\columnwidth}{0.4pt}\\[2pt]
        Correspondence to: \href{mailto:yileishao@sem.ecnu.edu.cn}{yileishao@sem.ecnu.edu.cn}
    }
}

\title{Is AI Widening the Wage Gap? A Hybrid Agentic Simulation for Labor Equity}

\author{Zhongbo Hu}
\affiliation{
  \institution{Shanghai AI-Finance School, East China Normal University}
  \city{Shanghai}
  \country{ China }}

\author{Zonghan Wu}
\affiliation{
  \institution{Shanghai AI-Finance School, East China Normal University}
  \city{Shanghai}
  \country{ China }}

\author{Georgina Curto}
\affiliation{
  \institution{United Nations University Institute in Macau}
  \city{Macau}
  \country{China}}

\author{Aocheng Tang}
\affiliation{
  \institution{Shanghai AI-Finance School, East China Normal University}
  \city{Shanghai}
  \country{ China }}

\author{Yilei Shao}
\affiliation{
  \institution{Shanghai AI-Finance School, East China Normal University}
  \city{Shanghai}
  \country{ China }}

\begin{abstract}
Artificial intelligence (AI) is reshaping labor markets, yet its effects on wage distribution remain insufficiently understood. Conventional analytical approaches are limited in their ability to directly examine the dynamic evolution of worker behavior and income distribution under sustained AI shocks and counterfactual policy scenarios. To address this limitation, we present a hybrid agentic framework, which aims to overcome the challenges of scalability and replicability of rule-based models and the limited explainability in LLM-agentic frameworks. Using a hybrid approach and sociodemographic data from China, we simulate and interpret changes in wage distribution under repeated AI shocks in the labor market. The results show that both the average-wage ratio between workers in the top and bottom income deciles (T10/B10) and the Gini coefficient increase persistently, suggesting that AI shocks widen the wage gap and exacerbate income inequality. This pattern of a widening wage gap remains robust across alternative large language model decision engines and 30 Monte Carlo simulations. We further conduct counterfactual policy experiments. The results show that education subsidies targeted at low-income workers increase both the number of skill-upgrading attempts and the number of successful upgrades, with particularly pronounced improvements in the upskilling success probability of workers in the bottom income decile. These effects enable the policy to partially mitigate wage inequality. The proposed framework provides an interpretable, scalable, and replicable simulation approach for examining the impact of AI shocks on wage distribution. It also offers policymakers a complementary analytical tool for evaluating policy interventions.
\end{abstract}


\newcommand{\BibTeX}{\rm B\kern-.05em{\sc i\kern-.025em b}\kern-.08em\TeX}

\begin{document}




\maketitle 
\thispagestyle{firstpage}
\pagestyle{empty}
\section{Introduction}\label{sec1}

With the advent of Large Language Models (LLMs), Artificial Intelligence (AI) has become pervasively integrated across diverse socio-economic sectors. The involvement of AI has significantly enhanced labor productivity, prompting producers to adopt AI technologies on an extensive scale. Although the impact of AI varies across different fields and industries, the potential risk of increased unemployment remains a major concern for the workforce\citep{nguyen2026ai}. Existing research suggests that in the current labor market, the augmentation effect of AI on occupations outweighs its substitution effect\citep{acemoglu2026building,bonney2024impact}. Nevertheless, it is undeniable that the application of automation technologies, such as AI-assisted office tools, has displaced numerous job positions. More granular studies indicate that AI exerts both substitution and complementary effects on job roles. In positions characterized by task substitution, specifically, the wage growth rate of workers is systematically lower than in positions where AI plays a complementary role \citep{acemoglu2022tasks}. Despite diverse investigations into AI's impact on the labor market, few scholars have focused on the issue of the AI-induced wage gap. Furthermore, conventional macroeconomic indicators, such as unemployment and aggregate inflation, as well as traditional methods such as rule-based ABMs and regression analysis, may not fully capture the dynamic evolution of this phenomenon. Consequently, three critical questions arise. First, does sustained AI exposure widen the wage gap among workers. Second, if wage disparities widen, what policy instruments can partially mitigate this effect. Third, how can these dynamics be effectively modeled and characterized.

Agentic frameworks provide a suitable environment for evaluating the questions outlined above and constitute a valuable methodological approach for public policy research. 
In social simulation research, agent behavior is traditionally specified through explicitly programmed rules, giving rise to what are commonly referred to as agent-based models (ABMs) \citep{Dignum2021}. ABMs have a long tradition in computer science \citep{retzlaff2021history} and are predominantly rule-driven or bounded-rational decision rules to govern
agents' behavior and interactions\citep{datseris2022agents,poledna2023economic,liu2024simulating}. Such rules may characterize the needs and motivations of individual agents \citep{Aguilera2024} or represent external environmental conditions, such as geographical settings \citep{Robinson2007Comparison} or policy environment\citep{Dweck2020Discussing}. As \citet{MacalNorth2005} suggests, incorporating multiple conditions and behavioral rules is often necessary to capture the growing complexity of real-world social environments. For example, labor-market simulations may require rules governing firms’ hiring and firing decisions, workers’ job search, worker–firm matching, and wage determination. 

By simulating social interactions and behavioral responses, ABMs provide a framework for assessing the impact of environmental change and policy interventions \citep{Bonabeau2002Agent}. One of their principal strengths lies in their transparency, as researchers can directly observe the rules governing agent behavior and trace the processes through which aggregate outcomes emerge. Nevertheless, conventional ABMs also face important limitations. In such models, agent behavior is largely determined by predefined rules, which can limit behavioral flexibility and adaptability as the complexity of the environment and the range of possible behaviors increase. Predetermined rules may not adequately capture the evolving and context-dependent nature of social environments, and models developed for one institutional or cultural setting may not be readily transferable to another.

The emergence of large language models (LLMs) has partially alleviated these limitations and stimulated the development of a wide range of LLM-based agentic frameworks. Trained on massive textual corpora, LLMs exhibit certain human-like language capabilities and can generate plausible context-sensitive responses to changing and complex environmental conditions \citep{Wei2022Emergent}. LLM-powered agents can therefore represent human-like social interactions and behaviors \citep{Piao2025AgentSociety}, offering new technical possibilities for social science research and policy analysis. For example, \citet{Ashery2025Emergent} leveraged interactions among LLM-powered agents to examine the emergence of collective social phenomena. Their experiments show that repeated local interactions can lead agent populations to spontaneously develop shared social conventions, collective biases, and tipping-point dynamics. However, the stand-alone use of LLMs also entails substantial limitations, including hallucinated decisions \citep{adel2025generative}, logical inconsistency \citep{cheng2025empowering}, and limited transparency and reproducibility \citep{sapkota2025comprehensive}.

Although a growing body of research has examined LLM-powered agents in social contexts \citep{Park2023,Li2024EconAgent}, both conventional rule-based ABMs and stand-alone LLM-based models exhibit significant shortcomings. Meanwhile, wage determination and income distribution in labor economics constitute another important area in which agentic approaches can generate valuable insights. To address these limitations, we develop a hybrid framework that combines explicit rules with LLM-based decision-making and apply it to investigate how AI shocks affect the wage gap. Within our framework, explicit rules and LLM-based decision-making play distinct but complementary roles. Specifically, at the aggregate level, the labor-market environment is governed by a transparent rule-based structure, whereas individual workers are represented by heterogeneous agents whose behavioral decisions are generated by an LLM. By structuring agent attributes and states—including age, occupation, and learning capacity—the framework enables LLM-powered agents to exhibit behavioral heterogeneity and respond to qualitative and context-dependent conditions in ways that are difficult to reproduce using predetermined rules alone. At the same time, the explicit rules governing the broader labor-market environment enhance the interpretability and traceability of an otherwise opaque LLM-agent system, particularly in simulations involving large-scale interactions among agents \citep{wang2024survey}.

The contributions of this research are threefold: (1) It presents a hybrid agentic methodology that integrates rule-based and LLM-based approaches, enhancing the scalability of rule-based models and the explainability of LLM-based models. (2) Relying on this framework, we offer empirical evidence that continuous AI shocks induce asymmetric wage distributional effects. (3) We utilize this framework to evaluate education subsidy policy interventions aimed at mitigating AI-induced wage polarization. By tracking how these interventions influence individual agent behavior and aggregate wage distribution, we provide policymakers with an effective analytical tool to evaluate policies that aim to mitigate inequity in the AI era.

\section{ Methodology}\label{sec:method}

\begin{figure*}[h]
    \centering
    \includegraphics[
    width=0.7\textwidth
]{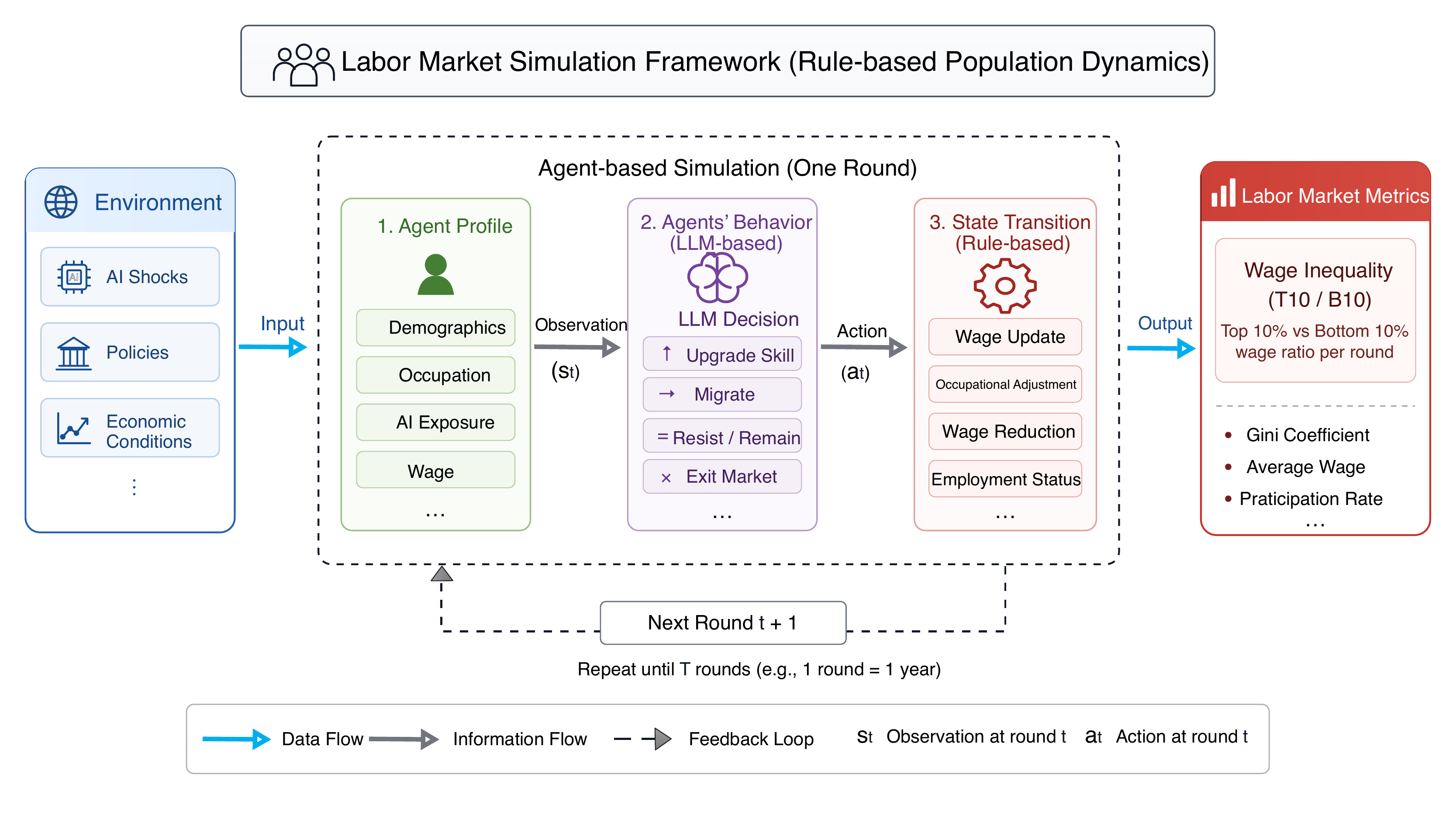}
    \caption{This figure illustrates the overall simulation pipeline. Under external conditions, including AI shocks, policies, and economic conditions, the LLM serves as the decision-making engine, generating decisions such as skill upgrading and occupational migration based on each agent’s individual state (e.g., demographics and occupation) and the external environment. Rule-based state-transition mechanisms then update wages, occupations, and employment status. Finally, labor-market outcomes, including the wage inequality measure (T10/B10) and the Gini coefficient, are calculated. Repeated simulation rounds are conducted to capture the dynamic evolution of wage distribution under AI shocks.}
    \label{fig:Pipeline}
\end{figure*}

\subsection{Overall Hybrid Simulation Model}\label{sec:env}
AI-driven wage gap is a distributional problem that unfolds through individual-level labor-market decisions, aggregated across heterogeneous workers. Studying it requires a model that can (i)~represent the overall labor market and individual behavior, (ii)~represent worker heterogeneity along multiple dimensions, (iii)~generate qualitative strategic choices that respond to each worker's specific circumstances, and (iv)~produce outcomes measured with the T10/B10 wage ratio (Ratio) to evaluate the wage gap. This section describes the methodology we use to meet these requirements with a hybrid agentic framework in which a rule-based labor-market environment is coupled with an LLM-driven individual agent's decision engine. The integrated simulation pipeline is illustrated in Figure~\ref{fig:Pipeline}. The agents represent the different profiles of workers, according to their demographics. Environmental factors (including AI shocks and policies to mitigate wage polarization) influence the LLM-ruled behavior of agents, towards the status of "upgrade", "migrate", "resist", or "exit" the labor market. The outcome of the agents' behavior change is measured with a wage inequality (T10/B10) ratio indicator.

Overall, our simulation proceeds in three stages. First, we construct the initial agent population by assigning each agent a set of sociodemographic and labor-market attributes. Here, initialization refers to assigning agents their starting sociodemographic and labor-market attributes at t=0 (Round 0). Some attributes, such as age cohort and adaptive elasticity, remain fixed throughout the simulation, whereas others, such as wages and employment status, evolve dynamically. Aggregate indicators, including the unemployment rate and the Gini coefficient, are subsequently calculated from the evolving agent-level states. Initialization takes place before the simulation formally begins and therefore involves no LLM-based decision-making, AI shocks, or subsequent changes in agents’ behavioral states and aggregate labor-market outcomes. Second, once the simulation begins, the LLM-based agents make decisions based on the environmental information available to them, leading to dynamic changes in their employment states and wages. GLM-4-Flash serves as the primary LLM decision engine, while Llama 3.1, Qwen3, and DeepSeek-V4 (DS) are used as alternative decision engines to assess the robustness of the simulation results. Third, the wage outcomes of all agents are aggregated to calculate macro-level indicators, including wage ratios and the Gini coefficient.

At the beginning of the simulation, we initialize 1,000 agents using sociodemographic data for China, primarily obtained from the National Bureau of Statistics of China (detailed descriptions of the data sources are provided in \textit{supplementary material}). The sociodemographic data covers aggregate labor market conditions, demographic composition, occupation-specific wages, and the distribution of AI exposure across occupations. Although our model is empirically grounded in real-world data, it may still be subject to certain biases in its initialization. To assess how closely the initialized model and data reproduce the real-world data, we compare the model-initiated outputs with their corresponding sociodemographic data. The results indicate that the initialized model closely matches the sociodemographic data: the absolute deviations for all proportion-based indicators are within 5 percentage points, while the relative deviations for all wage-related indicators are below 5\%, with a maximum deviation of 4.7\%. Detailed comparison results are reported in Appendix~\ref {app:Prompt}.

Furthermore, we formulate the simulation as a discrete-time system $\mathcal{M}$, defined as follows:
\begin{equation}
   \mathcal{M} = \langle \mathcal{N}, \mathcal{E}, \mathcal{S}, \mathcal{A}, f, T \rangle, 
\end{equation}
 where $\mathcal{N} = \{1, 2, 3, \dots\, N \}$ is the set of worker agents, and we set the number of agents to N=1000. $\mathcal{E}$ denotes the environmental state, which comprises the aggregate unemployment rate and the current simulation round; $\mathcal{S} = \{s_{i,t}\}_{i=1}^{N}$ denotes the collection of individual agent states and attributes at simulation round t. $\mathcal{A}$ denotes the action space, which includes occupational transitions and employment decisions, with each occupation associated with a corresponding level of AI exposure. $f\colon \mathcal{S} \times  \mathcal{E} \times \mathcal{A} \to \mathcal{S}$ is the state transition function governed by the LLM, which functions as the agents' behavioral engine to determine state updates based on previous and current conditions and environmental shocks(AI shock and policy intervention). $T$ is the simulation horizon in rounds. One round in the simulation is defined as a year in real-world time. The above definition also indicates that our model can be further extended, for example, by incorporating firm and government agents to form a multi-agent model that captures the interactions among different labor-market actors and their resulting effects
 
\subsection{Agentic demographics}
In the labor market, different types of occupations are associated with corresponding wage levels. In our framework, \textit{Routine} refers to occupations characterized by standardized, rule-based, and repetitive tasks. \textit{Manual} refers to occupations involving non-routine physical tasks that typically require physical presence, spatial judgment, sensory perception, or physical skills. \textit{Cognitive} refers to occupations involving non-routine cognitive tasks that rely on specialized knowledge, reasoning, creativity, analysis, and judgment. \textit{Managerial} refers to occupations centered on managerial decision-making, resource allocation, cross-team coordination, and strategic judgment. In general, workers in \textit{Managerial} occupations tend to earn higher wages than those in \textit{Routine} occupations. With the rapid development of AI, technological advances may affect workers' wages by displacing or transforming existing jobs. Workers facing such technological displacement may exit employment or transition to other occupations, thereby generating fluctuations in their earnings. To capture these labor-market dynamics, we model a single labor market with four occupational types---Routine, Manual, Cognitive, and Managerial---and three levels of AI exposure (Low, Medium, High), yielding twelve occupation-exposure cells. There is no product market or financial market. Accordingly, our model does not explicitly model the interaction between labor supply and labor demand; instead, it focuses on the effects of AI shocks on wage inequality among workers.

Although our analysis focuses exclusively on the labor market, it remains a highly complex system. To more closely represent labor-market dynamics under AI shocks and the states of individual agents, we assign each agent a set of sociodemographic attributes. Each agent~$i$ is initialized by sampling the attributes $g_i$, $o_i$, $a_i$, $e_i$, $w_i$, $\rho_i$, and $r_i$ from the corresponding distributions reported in the \textit{supplementary material.}Each agent~$i$ is characterized by a state vector $s_i = (g_i, a_i, o_i, e_i,w_i, \rho_i, r_i, h_i)$, where $g_i$ denotes the age cohort; $a_i$ denotes the AI exposure level, reflecting the extent to which the agent's occupation is exposed to AI-related technological change.$o_i$ denotes the occupation; $e_i \in \{0,1\}$ denotes the employment indicator; $w_i$ denotes the annual wage measured in Chinese yuan (RMB) (See Section \ref{Salary} for details); $\rho_i$ denotes the adaptive elasticity reflecting the ability to learn new skills and adapt to changes in the work environment. $r_i \in \{0,1\}$ denotes retirement eligibility and is specified as a binary parameter applicable only to agents in the 50--59 age cohort. This specification reflects gender differences in China's statutory retirement ages. Because our simulation does not explicitly distinguish between male and female agents, we do not assign a uniform retirement age to all agents. Instead, we allow retirement to occur within the 50--59 age cohort to approximate the heterogeneity in retirement timing between male and female workers. $h_i$ denotes the action-history buffer. The attributes $g_i$, $\rho_i$, and $r_i$ are assigned at initialization and remain fixed throughout the simulation, as they represent agents' initial endowments, including age, adaptive elasticity, and retirement probability. In contrast, $a_i$, $e_i$, $w_i$, and $h_i$ evolve endogenously over the course of the simulation in response to agents' actions. Since an agent's state is determined by multiple parameters, each agent is inherently heterogeneous.

\textit{Employment status.}
In real-world labor markets, re-employment following job loss typically involves job search, job matching, and skill adjustment, resulting in a time lag between unemployment and re-employment. Given that this study focuses on the dynamic evolution of wage inequality under AI shocks rather than short-term employment matching, we do not consider intra-round employment transitions. Specifically, agents who lose their jobs in a given year cannot be re-employed within the same year. At the same time, the initial employment status of agents needs to reflect real-world labor-market conditions. According to data published by the World Bank \citep{worldbank_china_data}, China's urban unemployment rate was 4.6\% in 2025. Accordingly, at $t=0$ (Round 0), we use this 4.6\% rate as the baseline parameter to determine the number of agents initialized as unemployed. In subsequent simulation rounds, a within-industry layoff rate ($\lambda$) of 3\% is applied in each round \citep{bls_jolts}. Data from the National Bureau of Statistics of China (NBS) indicate that young workers face a substantially higher unemployment rate. To capture this age-related disparity in unemployment risk, we set the initial unemployment probability for agents aged 20--29 to three times the baseline rate. This adjustment is designed to reflect the disproportionate challenges faced by youth in the Chinese labor market. 

In addition, when jobs are exposed to technological shocks and face the threat of displacement, workers may respond by switching occupations or remaining in their current positions at lower wages\citep{FossenSorgner2022}. This adjustment process may give rise to frictional unemployment. Furthermore,\citet{Wang2023Future} estimate that approximately 54\% of jobs in China could be at risk of substitution by AI over the coming decades. This suggests that workers affected by AI shocks may either leave employment or switch jobs. Drawing on their estimate, we introduce a model rule under which agents in occupations with high AI exposure have a 54\% probability of choosing to Migrate or Exit and a 46\% probability of choosing to Resist—that is, to remain in their current positions. Agents who choose to Resist incur a 25\% wage reduction\citep{Illing2024Gender}.

\textit{Wage.}\label{Salary} 
The determination of labor income, a central focus of this study, is influenced by a multitude of factors, including education, government policy, and seniority. An initial wage distribution that reflects real-world conditions is critical to ensuring the realism and validity of the simulation outcomes. To incorporate real-world-like characteristics of wage dispersion, we follow the Mincer\citep{Mincer1958Investment} tradition by assuming that initial annual wages, $w_i$, follow a log-normal distribution. The specification is as follows:
\begin{equation}
w_i \sim \text{LogNormal}\!\left(\mu_{o_i},\ \sigma = 0.30\right), \quad \mathbb{E}[w_i] = \bar{w}_{o_i}, \tag{2}
\end{equation}
where the mean annual wage for each occupation, $\bar{w}_{o_i}$, is calibrated using 2024 data from the NBS for urban employees: ¥78,561 for \textit{Routine}, ¥77,584 for \textit{Manual}, ¥148,046 for \textit{Cognitive}, and ¥203,014 for \textit{Managerial} roles. To account for intra-occupational heterogeneity, we set the dispersion coefficient to $\sigma = 0.30$\citep{erosa2024wage}, consistent with observed variations in individual labor income within the same sectors. Furthermore, to reflect institutional protections for workers' basic livelihood and rights, a statutory minimum wage floor is imposed. Based on the 2024 national average monthly minimum wage (¥2,100), we set the annual wage floor at $s_{\min} = \text{¥}25,200$. The remaining parameter configurations and their respective initialization values are detailed in the appendix~\ref{app:results}.

In addition to their regular wages, workers can earn a wage premium by acquiring advanced skills. For workers, AI technology therefore represents both a potential adverse shock and a source of economic gains. According to research by scholars at the Oxford Internet Institute and the Center for Social Data Science at the University of Copenhagen, workers who have mastered AI skills earn wages approximately 21\% above the average\citep{Oxford2023AISkills}. Consequently, in this study, when an agent successfully executes an Upgrade, its wage is updated as: $w_1 = w_0 \times 1.21$. Furthermore, according to opportunity cost theory, every economic activity entails an opportunity cost. When employed individuals devote time to learning in order to upgrade their skills, they forgo the income that could have been earned by allocating that time to alternative activities\citep{BianchiGiorcelli2022}. Accordingly, this study assumes that an employed agent who fails to complete a skill upgrade incurs a cost equivalent to 5\% of their annual wage.

\subsection{LLM-driven individual agents' (workers') behavior}
We introduce LLMs to drive the individual agents' decision-making process within the rule-based labor market behavioral space. The rationale for introducing the LLM as the decision-making engine is that it can adapt its decisions to changes in the environment\citep{gao2024large}, while allowing us to observe the resulting decisions in each round of the simulation. To enable the LLM to better perceive the employment environment and evaluate strategic trade-offs, we have developed prompts that translate quantitative parameters into qualitative narratives, as follows. 

\begin{enumerate}

\item \textbf{AI Exposure ($a_i$):}
Medium exposure level: ``You've started to feel AI's impact on your work.
Some colleagues' positions have been optimized away.
You're worried but not yet panicking.''

\item \textbf{Occupation ($o_i$):}
Routine occupation: ``You work on assembly lines or in standardized
operations---your tasks are highly repetitive.''

\item \textbf{Role assignment sets the agent's persona.}
You are a worker living in China. [Age narrative.] [Occupation narrative.]
Your current annual salary is RMB~[wage]. [AI-exposure narrative.]

\item \textbf{Behavioral options for individual agents.}
An employed agent replies with only one word:
\textsc{Resist}, \textsc{Migrate}, \textsc{Upgrade}, or \textsc{Exit}.
An unemployed agent replies with only one word:
\textsc{Seek} or \textsc{Remain}.
Thus, each agent selects an action according to the agent's current
employment state. The complete prompt is provided in the appendix \ref{app:Prompt}.

\end{enumerate}

In our simulation framework, agents’ decisions are generated by the LLM based on the current state of the environment and the agent's own action history. Agents face different sets of available actions depending on whether they are employed or unemployed. 
The action space for employed agents, $\mathcal{A}^{\text{emp}}$, is defined as follows:
\[
\mathcal{A}^{\text{emp}} =
\{\textsc{Upgrade}, \textsc{Migrate}, \textsc{Resist}, \textsc{Exit}\}.
\]
The action space for unemployed agents,$\mathcal{A}^{\text{unemp}}$, is defined as follows:
\[
\mathcal{A}^{\text{unemp}} =
\{\textsc{Seek}, \textsc{Remain}\}
\]
\textsc{Migrate} and \textsc{Seek} actions trigger a second-stage choice of a target occupation-exposure cell $(o', a')$. 
If the agent selects \textsc{Seek}, the agent may transition into any of the twelve occupational–exposure combinations in the subsequent round. 
When an agent selects the \textsc{Migrate} strategy, it may choose from nine alternative occupational–exposure combinations (i.e., transitioning from the current industry to another). In turn, when an agent selects the \textsc{Upgrade} strategy, it is deemed to have acquired AI-complementary skills, resulting in enhanced labor market competitiveness and elevated wages. If an agent selects \textsc{Exit}, it leaves employment and becomes unemployed. Further details on the agents’ decision-making process are provided in the appendix \ref{fig:LD}.

\section{Results}
\subsection{The Impact of AI-shocks in Wage Distribution}

\begin{table}[t]
\caption{Gini coefficients under the GLM scenario.}
\label{tab:gini-glm}
\centering
\small
\begin{tabular}{ccc}
\toprule
\textbf{Simulation Round}
& \textbf{All}
& \textbf{Employed Only} \\
\midrule
0 & 0.325 & 0.278 \\
1 & 0.339 & 0.290 \\
2 & 0.358 & 0.301 \\
3 & 0.364 & 0.311 \\
4 & 0.377 & 0.324 \\
5 & 0.388 & 0.339 \\
\bottomrule
\end{tabular}
\end{table}

\begin{figure}[t]
    \centering
    \includegraphics[
    width=0.5\textwidth
]{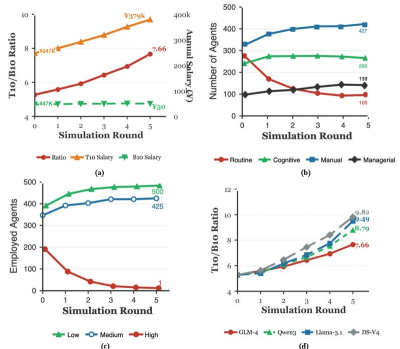}
    \caption{Multidimensional outputs of AI-driven wage polarization.
    (a) Evolution of wage polarization with and without AI exposure.
    (b) Changes in occupational composition.
    (c) Employment by AI-exposure level.
    (d) Comparison of results generated by different LLMs.}
    \label{fig:four-results}
\end{figure}

We first use the agentic framework to simulate the evolution of the wage distribution under sustained AI shocks in the Chinese labor market. The main results are in Figure~\ref{fig:four-results}. As shown in Figure~\ref{fig:four-results} (a), the average annual wage of agents in the top income decile (T10) increases from RMB~247K in Round~0 to RMB~379K in Round~5. By contrast, the average wage of workers in the bottom income decile (B10) rises only modestly, from RMB~47k to RMB~50k, representing substantially slower wage growth than that experienced by the T10 group. Consequently, the T10/B10 ratio(Ratio) increases steadily from 5.27 to 7.66.
Table~\ref{tab:gini-glm} further shows that the Gini coefficient increases from 0.325 to 0.388 from round 0 to round 5 of the simulation. The concurrent increases in the T10/B10 ratio and the Gini coefficient after the 5 simulation rounds indicate that sustained AI shocks are associated with greater wage polarization and rising income inequality in the simulated labor market.

The occupational distribution also changes substantially over the course of the simulation. Employment in \textit{Routine} occupations decreases from 276 to 100 workers, whereas employment in \textit{Manual}, \textit{Cognitive}, and \textit{Managerial} occupations increases from 326 to 427, from 236 to 260, and from 97 to 139 workers, respectively (Figure~\ref{fig:four-results}(b)). These results indicate a gradual reallocation of employment away from highly codified and repetitive \textit{Routine} occupations toward other occupational categories, particularly those associated with higher wages in the simulated environment. This pattern suggests that workers tend to transition from lower-wage to higher-wage occupations. Further decomposing employment by AI exposure reveals that the number of workers in the high-exposure group declines sharply from 180 to 1, while employment in the low- and medium-exposure groups increases from 408 to 500 and from 347 to 425, respectively(Figure~\ref{fig:four-results}(c)). This indicates that sustained AI shocks induce a pronounced shift in employment toward occupations with lower levels of AI exposure. One possible explanation is that workers in highly AI-exposed occupations face a greater risk of job displacement and therefore have stronger incentives to switch to less-exposed occupations~\citep{cazzaniga2025exposure}.The variation in employment across different levels of AI exposure demonstrates that using an LLM as the decision-making engine provides greater behavioral flexibility than a purely rule-based agent model.

\subsection{Sensitivity analysis of two key parameters}
\begin{figure}[t]
    \centering
    \includegraphics[
        width=0.5\textwidth,
    ]{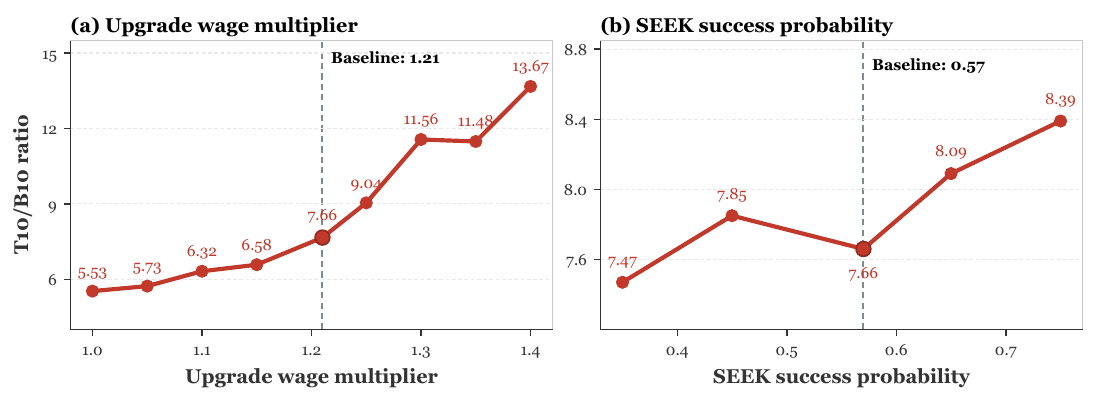}
    \caption{Sensitivity analysis of key parameters: (a) wage multiplier for successful skill upgrading; (b) probability of successful employment after choosing \textsc{Seek}.}
    \label{fig:sensitivity-test}
\end{figure}

Workers' earnings are closely related to wage premiums from skill upgrading and the likelihood of re-employment. Drawing on the skill wage premium literature \citep{KatzMurphy1992} and labor search-and-matching theory \citep{MortensenPissarides1994} as the theoretical foundations for these two processes, we conduct one-at-a-time sensitivity analyses on two key parameters: the wage multiplier associated with skill upgrading and the probability that an unemployed agent successfully finds employment after choosing \textit{Seek}. When varying each target parameter, all other model parameters are held constant to assess whether the baseline simulation results depend on specific parameter settings. The results are reported in Figure~\ref{fig:sensitivity-test}. Figure~\ref{fig:sensitivity-test}(a) shows that, although the Ratio exhibits small non-monotonic fluctuations across adjacent values of the upgrade wage multiplier, the overall relationship is strongly positive. This pattern is consistent with the model mechanism: a larger wage premium following successful skill upgrading produces greater wage gains for workers who complete the upgrading process, thereby disproportionately increasing wages at the upper end of the distribution. The Upgrade wage multiplier, therefore, affects the magnitude of wage polarization rather than its direction. Even when successful skill upgrading provides no additional wage premium, the Round~5 Ratio remains above its initial Round~0 value of 5.27.
Figure~\ref{fig:sensitivity-test}(b) indicates that the Ratio is less sensitive to the probability of successful employment after choosing \textit{Seek}. The response is not strictly monotonic around the baseline value: the Ratio is 7.85 when the success probability is 0.45 and 7.66 at the baseline value of 0.57. However, when the probability increases to 0.65 and 0.75, the ratio rises to 8.09 and 8.39, respectively. One possible explanation is that a higher re-employment probability increases employment without necessarily compressing the wage distribution, because newly employed agents may enter relatively low-wage occupations.

Overall, under every parameter setting considered, the Round~5 Ratio remains above its initial value. The qualitative conclusion that sustained AI shocks generate a wage gap between employed agents is therefore preserved.

\subsection{Robustness checks}
To assess the validity of the findings, we conduct two sets of robustness analyses: simulations using three alternative LLMs and Monte Carlo simulations. First, to assess whether the baseline results depend on a particular LLM-based decision engine, we conduct simulations using Qwen3, Llama-3.1, and DeepSeek-V4 (DS-V4), while holding the initialized population and all model parameters constant. As shown in Figure ~\ref{fig:four-results}(d), the T10/B10 ratio generated by GLM-4-Flash reaches 7.66 in Round~5, which is lower than the corresponding values produced by Qwen3 (8.78), Llama-3.1 (9.49), and DS-V4 (9.82). Although the magnitude of the wage gap varies across models, all LLMs exhibit a widening wage gap over the simulation horizon. This consistency indicates that the central finding of wage gap under AI shocks is not specific to GLM-4-Flash.

Second, we conduct 30 Monte Carlo simulations, each consisting of five rounds, and compare LLM-based and random decision-making under both AI-on and AI-off scenarios. The random-decision condition does not employ an LLM. Instead, each agent selects uniformly at random from its current feasible action set. Consequently, decisions in the random condition are independent of individual-level information, including the agent's age, occupation, wage, AI exposure, and adaptive capacity.

\begin{table}[t]
\caption{Monte Carlo Simulation Outputs}
\label{tab:anova}
\centering
\footnotesize
\setlength{\tabcolsep}{3pt}
\renewcommand{\arraystretch}{1.05}

\begin{tabular}{lccc}
\toprule
\textbf{State} &
\textbf{AI-On ($\pm$ SD)} &
\textbf{AI-Off ($\pm$ SD)} &
\textbf{Row Mean} \\
\midrule
LLM      & $7.310 \pm 0.411$ & $6.403 \pm 0.173$ & $6.857$ \\
Random   & $5.677 \pm 0.400$ & $5.765 \pm 0.346$ & $5.721$ \\
Col Mean & $6.494$           & $6.084$           & $6.289$ \\
\bottomrule
\end{tabular}
\end{table}

Table~\ref{tab:anova} shows that, regardless of whether AI shocks are present, the Ratio is consistently higher under LLM-based decision-making than under the corresponding random-decision condition. This result suggests that, relative to unconditional random behavior, heterogeneous decisions generated by LLMs on the basis of individual states further amplify disparities in the wage distribution. Moreover, the wage gap is stronger in the AI-on scenario than in the AI-off scenario, indicating that the model's AI-exposure mechanism and workers' decisions jointly contribute to the widening wage gap. These Monte Carlo results provide additional simulation-based robustness evidence for the baseline findings.

\subsection{Policy intervention}
To examine whether an education subsidy policy can mitigate AI-induced wage polarization, we introduce an education subsidy for AI-related skill development into the baseline simulation. We assume that the economy contains a dedicated funding pool that provides education subsidies to eligible workers in the Chinese context, whose annual income is below RMB~80,000. At the beginning of each simulation round, workers earning less than RMB~80,000 per year are informed of their eligibility for the subsidy before making labor market decisions. Eligible workers may then choose whether to participate in AI-related skills training. Participants receive a subsidy of RMB~3,000. The subsidy is recorded as a policy cost and is not directly included in workers' labor income.

Employed workers who participate in the training enter the skill-upgrading process. If the skill upgrade is successful, their wage is multiplied by 1.21; if it is unsuccessful, they incur the original 5\% training-failure cost. Unemployed participants do not incur any wage loss. Workers who are newly displaced during the current round become eligible to enter the education subsidy process only in the subsequent round.

\begin{figure*}[h]
    \centering
    \begin{tabular}{@{}ccc@{}}
        \includegraphics[width=0.31\textwidth]
        {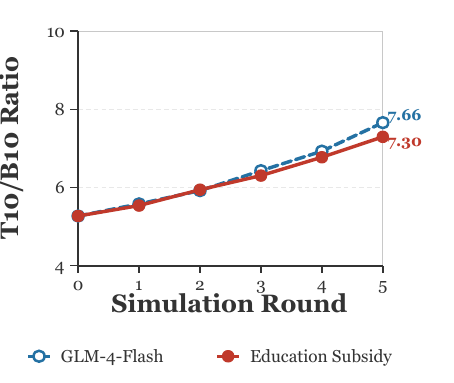} &
        \includegraphics[width=0.31\textwidth]
        {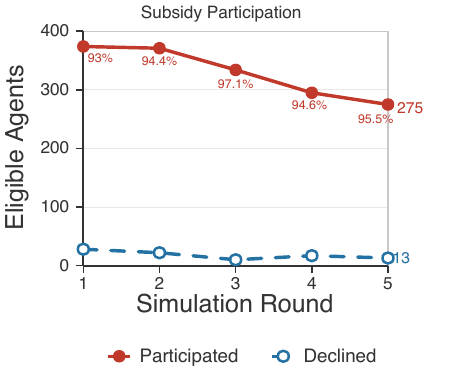} &
        \includegraphics[width=0.31\textwidth,]
        {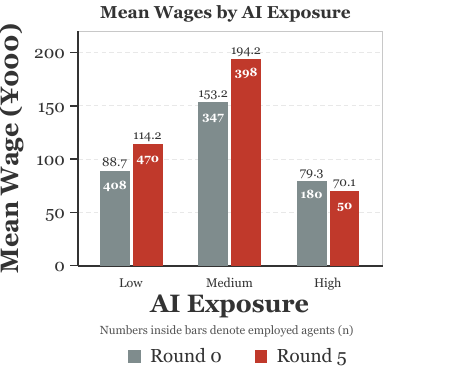} \\
        {\small (a)} & {\small (b)} & {\small (c)}
    \end{tabular}
    \caption{Effects and mechanisms of the education subsidy policy. 
    (a) presents the evolution of the T10/B10 ratio under the GLM baseline and education 
    subsidy policy scenarios; (b) shows subsidy participation; and (c) indicates mean wages 
    and employed-agent counts by AI-exposure level.}
    \label{fig:policy-results}
\end{figure*}

After the education subsidy policy is introduced into the simulation, the results show a gradual reduction in the wage gap relative to the baseline scenario. As illustrated in Figure~\ref{fig:policy-results}(a), the wage gap under the education subsidy policy remains identical to that in the baseline simulation during the first two rounds. From the third round onward, however, the wage gap under the policy intervention becomes smaller than that in the baseline scenario. These results suggest that the education subsidy policy can mitigate the wage gap to some extent, although its effects emerge only after a certain time lag \citep{AizerEtAl2024}.

To clarify the mechanisms through which the education subsidy policy operates, we examine two aspects: its incentive effect and its impact on the middle-income group. First, we analyze workers’ participation in subsidized training. Across the five simulation rounds, 93.0\%--97.1\% of eligible agents choose to participate in the training program in each round (Figure~\ref{fig:policy-results}(b)). We present a comparison of skill-upgrading activities between the baseline and education subsidy policy scenarios in Table~\ref{tab:upgrade_comparison}. Table~\ref{tab:upgrade_comparison} shows that relative to the baseline scenario, the number of skill-upgrading attempts increases from 3,289 to 3,836, while the number of successful upgrades rises from 1,673 to 1,807. These results indicate that the education subsidy effectively encourages workers to participate in training by increasing the total number of skill-upgrading attempts. However, the overall skill-upgrade success rate under the policy scenario is lower than that under the baseline scenario. This is because the policy primarily targets laborers with relatively low learning capacity, who consequently have a lower probability of successfully upgrading their skills. Nevertheless, a within-policy comparison between P-Round 1 and P-Round 5 shows that the skill-upgrade success rate increases over time. Taken together, the comparison between the policy and baseline scenarios suggests that the education subsidy policy encourages workers to participate in AI-related skill learning. However, the policy does not substantially improve the overall skill-upgrade success rate among employed workers. Within the policy scenario, however, the skill-upgrade success rate increases as the policy continues to be implemented. These findings suggest that the education subsidy policy primarily promotes skill development by increasing workers' participation in skill learning and gradually improving skill-upgrade outcomes over time, while its overall effectiveness remains constrained by the relatively low learning capacity of the targeted workers.

\begin{table}[t]
\caption{Comparison of Skill-Upgrading Outcomes Across Simulation Scenarios}
\label{tab:upgrade_comparison}
\centering
\small

\begin{tabular}{lcccc}
\toprule
\textbf{Indicator} &
\textbf{Baseline} &
\textbf{Policy} &
\textbf{P-Round1} &
\textbf{P-Round5} \\
\midrule
Upgrade attempts     & 3,289  & 3,836  & 85    & 77    \\
Successful upgrades  & 1,673  & 1,807  & 37    & 36    \\
Upgrade success rate & 50.9\% & 47.1\% & 43.5\% & 46.7\% \\
\bottomrule
\end{tabular}

\end{table}

We further examine workers located between the 25th and 75th percentiles of the income distribution(define it as the middle-income group) in each simulation round(Table~\ref{tab:middle_income_policy}). Employment in this middle-income group decreases slightly from 467 workers in Round~0 to 458 workers in Round~5, while its average annual wage increases from RMB~95,400 to RMB~118,200. The employment composition by AI exposure also shifts away from highly exposed occupations toward occupations with medium or low AI exposure(Figure~\ref{fig:policy-results}(c)). Nevertheless, the number of workers employed in high-AI-exposure occupations under the subsidy policy is approximately 50 times that observed in the no-policy scenario. This result suggests that the education subsidy enables some agents to acquire AI-related skills and remain in, or transition into, occupations with high levels of AI exposure.

\begin{table}[h]
\caption{Round-Specific P25--P75 Income Group Under the Education Subsidy Policy}
\label{tab:middle_income_policy}
\centering
\small
\begin{tabular}{lrrrr}
\toprule
& \multicolumn{2}{c}{\textbf{Round 0}} & \multicolumn{2}{c}{\textbf{Round 5}} \\
\cmidrule(lr){2-3}\cmidrule(lr){4-5}
\textbf{Group} & \textbf{Agents} & \textbf{Mean Wage} & \textbf{Agents} & \textbf{Mean Wage} \\
\midrule
P25--P75 & 467 & 95,400 & 458 & 118,200 \\
\bottomrule
\end{tabular}
\end{table}

Overall, the education subsidy increases training participation and the number of successful skill upgrades among simulated agents, while partially reducing the T10/B10 ratio. It also raises the average wage of the middle-income group. More importantly, the policy enhances workers’ capacity to adapt to a labor market increasingly affected by AI-related technological shocks.

\section{Discussion}
The purpose of this study is to examine whether and how AI shocks affect wage disparities among workers and to explore the policy measures that may be adopted in response to widening wage inequality. To address these questions, we develop a hybrid framework and conduct simulation experiments. 
The simulation results show that sustained AI shocks progressively widen wage disparities and increase the Gini coefficient. 
This process is closely associated with changes in occupational structure. 

AI shocks affect not only workers’ earnings but also their occupational choices and the overall structure of employment. When confronted with a high risk of technological substitution, workers in the simulation tend to leave highly exposed positions and move toward occupations that are less exposed or more adaptable to technological change. This finding is consistent with the central arguments of the skill-biased technological change and task-substitution literature, which emphasize that the benefits and adjustment costs associated with new technologies are not distributed evenly across workers\citep{acemoglu2002technical, autor2003skill}. Technological shocks can consequently induce structural adjustments in the labor market. Large-scale occupational mobility represents one manifestation of this adjustment process. 

From a governance perspective, the simulation allows to evaluate a diversity of policy frameworks. We incorporate an education subsidy policy in the Chinese labor market scenario. The policy evaluation within the simulated framework shows that subsidizing AI-related skills training for low-income workers can partially mitigate wage polarization. Relative to the baseline scenario, the wage gap narrows under the subsidy policy. This effect does not emerge immediately but becomes more pronounced after several simulation rounds, consistent with the widely recognized time lag in policy implementation and effectiveness\citep{card2018what，dawid2014economic}. Nevertheless, the overall policy effect remains limited. Although the subsidy reduces the wage gap and raises the average wage of middle-income workers, it does not produce a clear reduction in the Gini coefficient within the current simulation horizon. Education subsidies should therefore be understood as a policy instrument for mitigating wage polarization and strengthening workers’ capacity to adapt to technological change, rather than as a stand-alone solution to income inequality. 

The hybrid agentic framework proposed in this study integrates the stability of rule-based mechanisms with the autonomous decision-making capabilities of LLMs, enabling simulated agents to adapt their behavior dynamically in response to changes in the environment. In this way, the framework retains the interpretability of traditional ABMs while improving behavioral flexibility and model scalability. Nevertheless, as with other studies of complex social simulation, the present study is subject to several limitations that warrant further investigation in future research. First and foremost, the findings are generated within a simulated and simplified labor market. Therefore, results should be interpreted as mechanism-based evidence rather than validated forecasts of China’s actual wage distribution. 
Second, worker decisions are informed by LLMs. These models can generate context-dependent behavioral heterogeneity, but their responses may vary across model versions, prompt designs, and repeated API calls. The comparison across alternative LLMs and the parameter-sensitivity analysis reduce, but do not eliminate, this source of uncertainty.
Third, the current framework models the worker side of the labor market without explicitly representing firms, vacancies, labor demand, productivity, or endogenous wage bargaining. Wages therefore evolve according to predefined transition rules rather than through a market-clearing process.

Building on the limitations discussed above, future research could extend the current framework from the perspectives of disaggregated data comprehensiveness, market structure complexity, and policy evaluation preparedness. While the present study relies primarily on aggregate statistics, which limits its ability to explain heterogeneity in individual behavior, subsequent research could incorporate individual-level demographic, psychological, and behavioral data, expand the number of agents, and allow communication and interaction among them. These extensions would enable the model to more realistically reproduce worker behavior and social relationships.
Finally, the framework could be further extended by incorporating a dynamic policy environment. The government could adjust education subsidies, vocational training programs, unemployment protection, and incentives for firms’ technology adoption in response to changes in employment, wage inequality, and the skill composition of the workforce. This extension would facilitate a more systematic analysis of the temporal effects, interactions, and long-term welfare implications of alternative policy combinations.
\section{Conclusion}
This study develops a hybrid agentic framework to examine how AI shocks affect wage polarization and distributional inequity in China. The simulation results provide evidence that both the T10/B10 ratio and the Gini coefficient increase persistently under repeated AI shocks, indicating a widening wage gap and rising income inequality in the simulated labor market. Robustness tests using multiple large language models (LLMs) and Monte Carlo simulations preserve the direction of these trends, suggesting that the main findings are not specific to a particular decision engine or simulation run. We further conduct an evaluation of an education subsidy policy within the simulated labor market framework. The policy experiment illustrates how the T10/B10 income ratio declines when training subsidies encourage low-income workers to acquire AI-related skills. This finding suggests that subsidized education and skills training could serve as an important policy instrument for mitigating AI-induced wage polarization by strengthening workers’ capacity to adapt to technological change. Beyond its substantive findings, the proposed framework offers a methodological contribution and a practical analytical tool for policy evaluation in the labor market within the AI era.

\clearpage
\onecolumn
\appendix

\section{Model Calibration Results}
\label{app:results}
\begin{table*}[!ht]

\caption{Calibration of the Initial Agent Population: Demographic Composition}
\label{tab:initial-calibration-demographic}
\centering
\small
\setlength{\tabcolsep}{3pt}
\renewcommand{\arraystretch}{1.0}

\begin{tabular}{p{3.15cm}p{2.55cm}p{2.65cm}p{1.35cm}p{3.75cm}}
\toprule
\multicolumn{1}{c}{\textbf{Indicator}} &
\multicolumn{1}{c}{\textbf{SD}} &
\multicolumn{1}{c}{\textbf{Initialization}} &
\multicolumn{1}{c}{\textbf{Difference}} &
\multicolumn{1}{c}{\textbf{Source}} \\
\midrule
Age share($g_i$): 20--29 (\%) & 22.0\% & 21.7\% & -0.3 pp & NBS, Seventh National Population Census (2020) \\
Age share($g_i$): 30--39 (\%) & 25.0\% & 23.6\% & -1.4 pp & Same as above \\
Age share($g_i$): 40--49 (\%) & 28.0\% & 29.5\% & +1.5 pp & Same as above \\
Age share($g_i$): 50--59 (\%) & 25.0\% & 25.2\% & +0.2 pp & Same as above \\
Occupational share($o_i$): Routine (\%) & 30.0\% & 29.4\% & -0.6 pp & NBS 19-industry employment (2025) \\
Occupational share($o_i$): Manual (\%) & 35.0\% & 34.6\% & -0.4 pp & Same as above \\
Occupational share($o_i$): Cognitive (\%) & 25.0\% & 26.0\% & +1.0 pp & Same as above \\
Occupational share($o_i$): Managerial (\%) & 10.0\% & 10.0\% & 0.0 pp & Same as above \\
\bottomrule
\end{tabular}

\vspace{2pt}

\begin{minipage}{0.98\textwidth}
\footnotesize
\textit{Notes:} Data were obtained from the National Bureau of Statistics of China (NBS). 
SD denotes sociodemographic data. The initialization snapshot comprises 1,000 agents. 
Differences are calculated as the model-initialization values minus the corresponding SD values. 
The abbreviation ``pp'' denotes percentage points and is used for shares, unemployment rates, 
and AI-exposure probabilities, whereas percentage differences are reported for wages.
\end{minipage}

\end{table*}

\begin{table*}[h]
\caption{Calibration of the Initial Agent Population: Macroeconomic and Distributional Indicators}
\label{tab:initial-calibration-macro}
\centering
\small
\setlength{\tabcolsep}{1.6pt}
\renewcommand{\arraystretch}{0.92}
\begin{tabular}{@{}p{3.15cm}p{2.55cm}p{2.65cm}p{1.35cm}p{3.75cm}@{}}
\toprule
\multicolumn{1}{c}{\parbox{3.15cm}{\centering\textbf{Indicator}}} &
\multicolumn{1}{c}{\parbox{2.55cm}{\centering\textbf{SD}}} &
\multicolumn{1}{c}{\parbox{2.65cm}{\centering\textbf{Initialization}}} &
\multicolumn{1}{c}{\parbox{1.35cm}{\centering\textbf{Difference}}} &
\multicolumn{1}{c}{\parbox{3.75cm}{\centering\textbf{Source}}} \\
\midrule 
Initial unemployment rate($\lambda$): ages 20--29 (\%) & 13.8\% ($3 \times 4.6\%$) & 11.1\% & -2.7 pp & Calculated by author \\
Initial unemployment rate($\lambda$): ages 30--59 (\%) & 4.6\% & 5.2\% & +0.6 pp & World Bank (2025) \\
Annual minimum-wage floor(($w_i$)) (RMB) & 25,200 & 25,200 & 0.0\% & NBS 2024(¥2,100)× 12 \\
\bottomrule
\end{tabular}

\par\medskip
\noindent
\parbox{0.98\textwidth}{%
\footnotesize
\textit{Notes:} Same as Table~ ~\ref{tab:initial-calibration-demographic}. The all-agent unemployment target is implied by the age-specific unemployment rule and target age composition.}
\end{table*}

\begin{table*}[h]
\caption{Calibration of the Initial Agent Population: Occupation-Specific Annual Wages}
\label{tab:initial-calibration-wages}
\centering
\small
\setlength{\tabcolsep}{1.6pt}
\renewcommand{\arraystretch}{0.92}
\begin{tabular}{@{}p{3.15cm}p{2.55cm}p{2.65cm}p{1.35cm}p{3.75cm}@{}}
\toprule
\multicolumn{1}{c}{\parbox{3.15cm}{\centering\textbf{Indicator}}} &
\multicolumn{1}{c}{\parbox{2.55cm}{\centering\textbf{SD}}} &
\multicolumn{1}{c}{\parbox{2.65cm}{\centering\textbf{Initialization}}} &
\multicolumn{1}{c}{\parbox{1.35cm}{\centering\textbf{Difference}}} &
\multicolumn{1}{c}{\parbox{3.75cm}{\centering\textbf{Source}}} \\
\midrule
Routine wage($w_i$), & ¥78,561 & ¥80,194 & +2.1\% & NBS Average Wage of Urban
Units (2024) \\
Manual wage($w_i$) & ¥77,584 & ¥78,639 & +1.4\% & Same as above \\
Cognitive wage($w_i$) & ¥148,046 & ¥149,275 & +0.8\% & Same as above \\
Managerial wage($w_i$) & ¥203,014 & ¥212,514 & +4.7\% & Same as above \\
\bottomrule
\end{tabular}

\par\medskip
\noindent
\parbox{0.98\textwidth}{%
\footnotesize
\textit{Notes:} Same as Table~ ~\ref{tab:initial-calibration-demographic}. Occupation-specific initial wages are calculated among employed agents only.}
\end{table*}
\vspace{-1em}

\begin{table*}[!htbp]
\caption{Calibration of the Initial Agent Population: Occupation-Conditional AI Exposure Distribution}
\label{tab:initial-calibration-ai-exposure}
\small
\setlength{\tabcolsep}{2.2pt}
\renewcommand{\arraystretch}{0.92}
\begin{tabular}{@{}lccc ccc ccc p{2.8cm}@{}}
\toprule
\textbf{Occupation} &
\multicolumn{3}{c}{\textbf{SD (\%)}} &
\multicolumn{3}{c}{\textbf{Initialization (\%)}} &
\multicolumn{3}{c}{\textbf{Difference (pp)}} &
\textbf{Source} \\
\cmidrule(lr){2-4}\cmidrule(lr){5-7}\cmidrule(lr){8-10}
& L & M & H & L & M & H & L & M & H & \\
\midrule
Routine($a_i$) & 5.2 & 33.1 & 61.7 & 6.5 & 29.6 & 63.9 & +1.3 & -3.5 & +2.2 & ChinaJob exposure index\citep{hao2026chinajob} \\
Manual($a_i$) & 100.0 & 0.0 & 0.0 & 100.0 & 0.0 & 0.0 & 0.0 & 0.0 & 0.0 & Same as above \\
Cognitive($a_i$) & 31.4 & 66.4 & 2.2 & 29.2 & 69.2 & 1.5 & -2.2 & +2.8 & -0.7 & Same as above \\
Managerial($a_i$) & 0.0 & 100.0 & 0.0 & 0.0 & 100.0 & 0.0 & 0.0 & 0.0 & 0.0 & Same as above \\
\bottomrule
\end{tabular}

\par\medskip
\noindent
\parbox{0.98\textwidth}{%
\footnotesize
\textit{Notes:} Same as Table~ ~\ref{tab:initial-calibration-demographic}. L/M/H denotes low, medium, and high AI exposure, respectively. Each row reports $P(a_i\mid o_i)$ by occupation; each difference is calculated as the initialized share minus its occupation-specific target.}
\end{table*}

\clearpage
\section{the Complete Prompt}
\label{app:Prompt}
\begin{table*}[h]
\caption{AI-exposure prompt components.}
\label{tab:prompt-ai-exposure}
\centering
\small
\begin{tabular}{@{}p{2.2cm}p{10.8cm}@{}}
\toprule
\textbf{AI exposure} & \textbf{Prompt} \\
\midrule
Low & Your job faces relatively little AI disruption for now; it feels safe in the short term. \\
Medium & You have started to feel AI's impact on your work. Some colleagues' positions have been optimized away. You are worried but not yet panicking. \\
High & Your job is being rapidly replaced by AI. People around you keep getting laid off. You feel anxious every day, unsure how much longer you can hold on. \\
\bottomrule
\end{tabular}
\end{table*}

\begin{table*}[h]
\caption{Occupation prompt components.}
\label{tab:prompt-occupation}
\centering
\small
\begin{tabular}{@{}p{2.2cm}p{10.8cm}@{}}
\toprule
\textbf{Occupation} & \textbf{Prompt} \\
\midrule
Routine & You work on assembly lines or in standardized operations; your tasks are highly repetitive. \\
Manual & You do physical labor or service work, such as logistics, food service, construction, and similar jobs. \\
Cognitive & You do knowledge work that requires professional expertise, such as programming, design, and analysis. \\
Managerial & You are in management, responsible for team leadership and business decisions. Your income is higher, but so is the pressure. \\
\bottomrule
\end{tabular}
\end{table*}

\clearpage
\section{LLM Decision-Making Process}
\label{fig:LD}
\begin{figure*}[h]
    \centering
    \includegraphics[
        width=0.65\textwidth
    ]{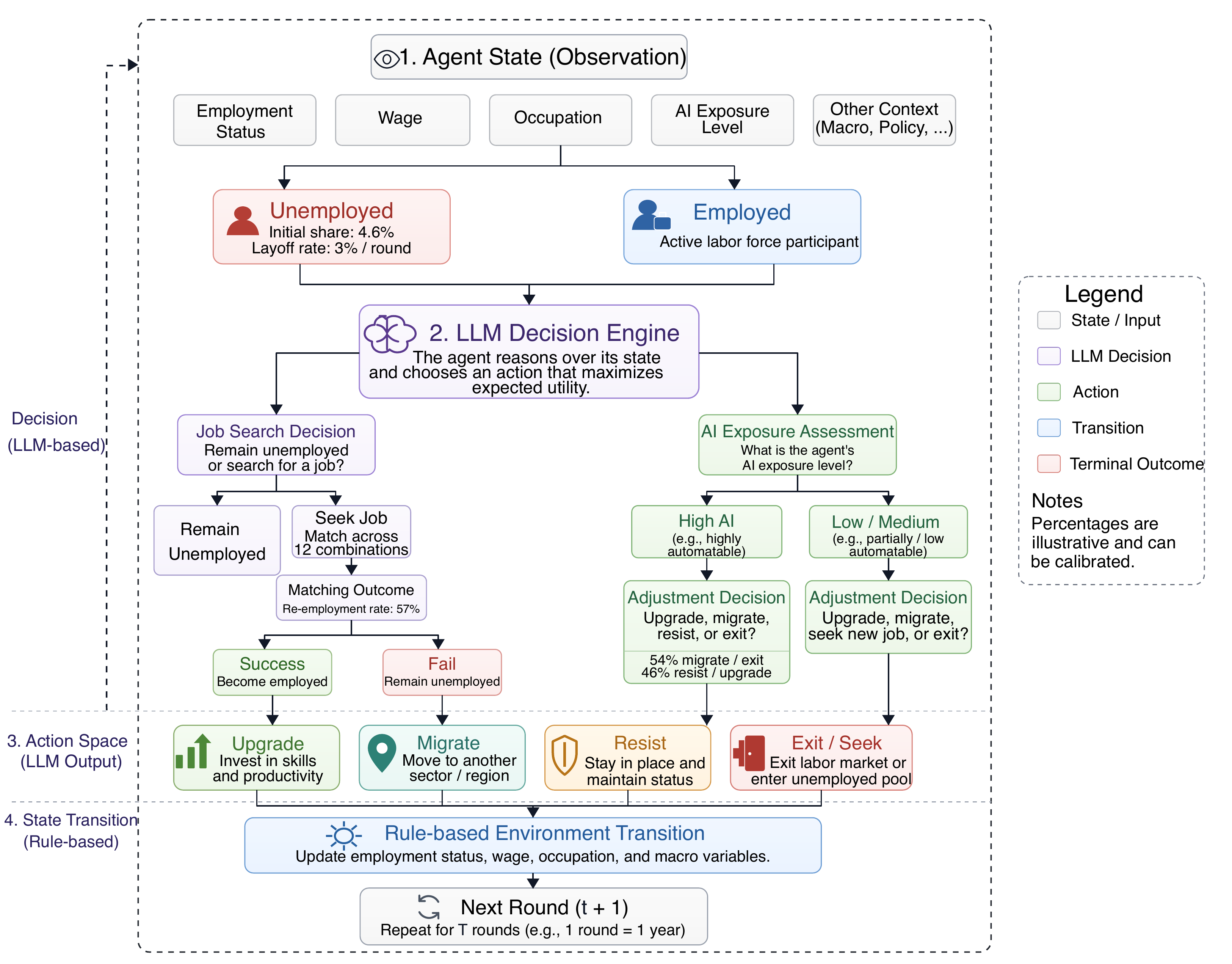}
    \caption{This figure illustrates the decision-making and state-transition process of worker agents in each simulation round, with the LLM serving as the decision-making engine. The LLM first observes each agent’s employment status, occupation, wage, AI exposure, and other relevant information, and then selects actions such as Upgrade, Migrate, Resist, and Exit/Seek from the action space according to the agent’s employment status. Subsequently, rule-based state-transition mechanisms update employment status, wages, and occupations before the simulation proceeds to the next round.}
    \label{fig:agent-decision}
\end{figure*}

\clearpage 
\twocolumn
\bibliographystyle{ACM-Reference-Format} 
\bibliography{sample}

\end{document}